\documentclass[nocompress,a4paper]{spie}  
\usepackage[]{graphicx}
\usepackage{amsmath} 
\newcommand{\angstrom}{\textup{\AA}}

\title{Status of detector development for the European XFEL} 
   
\author{Jolanta Sztuk-Dambietz\supit{a}, Steffen Hauf\supit{a}, Andreas Koch\supit{a}, Markus Kuster\supit{a}, Monica Turcato\supit{a}
\skiplinehalf
\supit{a}The European XFEL GmbH, Albert-Einstein-Ring 19, Hamburg, Germany; \\
} 
  
\authorinfo{Further author information: (Send correspondence to J. Sztuk-Dambietz)\\
            J. Sztuk-Dambietz.: E-mail: jolanta.sztuk@xfel.eu, Telephone: +49 40 8998 6925}

\begin{document} 
\maketitle 
 
\begin{abstract} 
The European X-ray Free Electron Laser (XFEL.EU) will provide as-yet-unrivaled peak
brilliance and ultra-short pulses of spatially coherent X-rays with a pulse length 
of less than 100 fs in the energy range between 0.25 and 25 keV. 
The high radiation intensity and ultra-short pulse duration will open a window for novel 
scientific techniques and will allow to explore new phenomena in biology, chemistry, material 
science, as well as matter at high energy density, atomic, ion and molecular physics.
The variety of scientific applications and especially the unique XFEL.EU time structure require 
adequate instrumentation to be developed in order to exploit the full potential of the light source. 
To make optimal use of the unprecedented capabilities of the European XFEL and master these vast 
technological challenges, the European XFEL GmbH has started a detector R\&D program.  
The technology concepts of the detector system presently under development are complementary in their
 performance and will cover the requirements of a large fraction of the scientific applications 
envisaged for the XFEL.EU facility. 
The actual status of the detector development projects which includes ultra-fast 2D imaging 
detectors, low repetition rate 2D detectors as well as strip detectors for e.g. spectroscopy applications 
and the infrastructure for the detectors' calibration and tests will be presented.
Furthermore, an overview of the forthcoming implementation phase of the European XFEL 
in terms of detector R\&D will be given.  
\end{abstract}



\section{Introduction}
\label{sec:intro}  

The European X-ray Free-Electron Laser (XFEL.EU) is a 3.4-kilometer-long international research facility currently 
under construction in the Hamburg, Germany area, that will start user operation in 2016~\cite{Altarelli06,Tschentscher10}. 
It is an X-ray photon source providing laterally coherent X-rays for six experimental stations 
(start-up configuration) in the range of approximately 250 eV to 25 keV. 
An electron beam is accelerated on a linear path up to energies of 17.5 GeV by superconducting cavities.
The electrons then generate coherent X-rays in a series of undulators up to 200 m in length based on 
the SASE process (Self-Amplified Spontaneous Emission). 
The electron acceleration, photon beam generation and beam transport path extends over a
total length of 3.4 km providing particularly intense and short X-ray pulses down to a few
femtoseconds with a peak brilliance of $10^{33}$ photons/s/mm$^2$/mrad$^2$/0.1\%BW. 
The electron source will deliver trains of typically 2700 pulses at a pulse repetition rate of 4.5~MHz. 
Each train will be followed by a gap of 99.4 ms.
The unique features of the European XFEL drive the need to develop dedicated instrumentation 
and detectors in particular. The instruments are optimized for particular purposes.
%
%
Each experiment requires light sources with special properties, such that the instruments are permanently 
assigned to the different X-ray sources (beamlines) of the European XFEL. There are six scientific instruments
planned to be installed in the start-up phase of the project. 
The list of the start-up instruments is shown in the Table~\ref{tab:experiments}.

\begin{table}[h] 
\caption{Scientific instruments at the XFEL.EU.} 
\label{tab:experiments}
\begin{center}       
\begin{tabular}{|p{0.25\linewidth}|p{0.55\linewidth}|} 
\hline 
{\bf Instrument} & {\bf Main application} \\
\hline
\multicolumn{2}{|c|}{{\bf SASE 1} $E \in$ 3-20~keV}\\ 
\hline
Single Particles, clusters, and Biomolecules (SPB)~\cite{SPB_CDR} & 
Structure determination of single particles - atomic clusters, biomolecules, virus particles, cells. \\
\hline
Femtosecond X-ray Experiments (FXE)~\cite{FXE_TDR} & 
Time-resolved structure investigations and dynamics investigation of solids, liquids, gases. \\
\hline
\multicolumn{2}{|c|}{{\bf SASE 2} $E \in$ 3-24(36)~keV} \\
\hline
Materials Imaging and Dynamics (MID)~\cite{MID_CDR} & 
Structure determination of nano-devices and dynamics at nano-scale \\
\hline
High-Energy Density matter experiments (HED)~\cite{HED_CDR} &
Investigation and excitations of matter under extreme conditions using hard X-ray FEL radiation, e.g. for probing dense plasma.\\
\hline
\multicolumn{2}{|c|}{{\bf SASE 3} $E \in$ 0.26-3~keV} \\
\hline
Small Quantum Systems (SQS)~\cite{SQS_TDR} &
Ultra-fast investigation (spectroscopy) of atoms, ions, molecules and clusters in intense fields and non-linear phenomena. \\
\hline
Spectroscopy \& Coherent Scattering (SCS)~\cite{SCS_CDR}&
Structure and dynamics of nano-systems and of non-reproducible biological objects.
Electronic and atomic structure as well as  dynamics of soft and hard matter, bio species and nanomaterials using soft X-rays.\\
\hline
\end{tabular}  
\end{center}   
\end{table}  
%

\section{Detectors at the XFEL.EU Scientific Instruments} 
\label{sec:challenges}
The different scientific applications at the XFEL.EU will require  different instrumentation,
in particular different imaging (2D) and spectroscopy (1D) detectors. Tables~\ref{tab:2d} and \ref{tab:1d} 
show the summary of the main detector requirements for all scientific instruments which are foreseen at the XFEL.EU.
\begin{table}[h]
\caption{Imaging detector requirements for the scientific instruments at the XFEL.EU.} 
\label{tab:2d}
\begin{center}       
\begin{tabular}{|p{0.15\linewidth}||p{0.11\linewidth}| p{0.11\linewidth}|p{0.11\linewidth}|p{0.11\linewidth}|p{0.11\linewidth}|p{0.11\linewidth}|} 
\hline
& {\bf SPB } & {\bf FXE} & {\bf MID} & {\bf HED} & {\bf SQS} & {\bf SCS}\\
\hline
{\bf Requirements} & \multicolumn{6}{|c|}{\bf Large Area Imaging Detectors}\\
\hline
\rule[-1ex]{0pt}{3.5ex} Repetition~rate & \multicolumn{6}{c|}{ up to 4.5 MHz} \\
\hline
\rule[-1ex]{0pt}{3.5ex} Sensitive energy range (Quantum Efficiency (QE) $>$80\%  & 3 - 16~keV & 6 - 25~keV &  3 - 25~keV & 4 - 25~keV & 0.26 - 3~keV & 0.26 - 3~keV \\
\hline
\rule[-1ex]{0pt}{3.5ex} Pixel~size & $\le$ 200$\mu m$ &$\le$500$\mu m$ &$\le$50$\mu m$ & $\le$100$\mu m$ &  $\le$ 200$\mu m$& $\le$ 200$\mu m$ \\
\hline
\rule[-1ex]{0pt}{3.5ex} Pixel~number &  \multicolumn{6}{c|}{1000 $\times$ 1000 or more}\\
\hline
\rule[-1ex]{0pt}{3.5ex} {Radiation hard design} & \multicolumn{6}{c|}{yes} \\
\hline
\rule[-1ex]{0pt}{3.5ex} Single photon sensitivity & \multicolumn{6}{c|}{yes} \\
\hline
\rule[-1ex]{0pt}{3.5ex} Dynamic range & $\ge$10$^{6}$ & 10$^5$ & $\ge$10$^3$ & $\ge$10$^3$& $\ge$10$^3$& $\ge$10$^3$\\
\hline
\rule[-1ex]{0pt}{3.5ex} Vacuum compatibility & 10$^{-3}$ mbar & gas/ambient pressure &  10$^{-3}$ mbar & \multicolumn{3}{c|}{{10$^{-6}$} mbar}\\
\hline
\rule[-1ex]{0pt}{3.5ex} Variable central hole & \multicolumn{6}{c|}{yes} \\
\hline
\rule[-1ex]{0pt}{3.5ex} VETO capability & \multicolumn{6}{c|}{yes} \\
\hline
\end{tabular} 
\end{center} 
\end{table} 
 
\begin{table}[h]
\caption{ Strip detector  requirements for the XFEL.EU scientific instruments at SASE 1 and SASE 2 (hard-X-rays).} 
\label{tab:1d}
\begin{center}       
\begin{tabular}{|p{0.15\linewidth}||p{0.11\linewidth}| p{0.11\linewidth}|p{0.11\linewidth}|p{0.11\linewidth}|} 
\hline
 & {\bf SPB } & {\bf FXE} & {\bf MID} & {\bf HED} \\
\hline
\rule[-1ex]{0pt}{3.5ex} {\bf Requirements} & \multicolumn{4}{|c|}{\bf Strip detectors for hard X-rays}\\
\hline
\rule[-1ex]{0pt}{3.5ex} Repetition~rate & \multicolumn{4}{c|}{ up to 4.5 MHz} \\
\hline
\rule[-1ex]{0pt}{3.5ex} Sensitive energy range & 3 - 16~keV & 6 - 25~keV & 6 - 25~keV & 4 - 25~keV     \\
\hline
\rule[-1ex]{0pt}{3.5ex} Strip pitch  & \multicolumn{4}{c|} {50 $\mu m$}      \\
\hline
\rule[-1ex]{0pt}{3.5ex} Strip number &  \multicolumn{4}{c|} {$\sim 1000$} \\
\hline
\rule[-1ex]{0pt}{3.5ex} Radiation hard design & \multicolumn{4}{c|} {no direct beam}  \\
\hline
\rule[-1ex]{0pt}{3.5ex} Single photon sensitivity & \multicolumn{4}{c|} {yes} \\
\hline
\rule[-1ex]{0pt}{3.5ex} Dynamic range & \multicolumn{4}{c|} {10$^4$}\\
\hline
\rule[-1ex]{0pt}{3.5ex} Vacuum compatibility &  \multicolumn{4}{c|} {10$^{-5}$ mbar} \\
\hline
\rule[-1ex]{0pt}{3.5ex} VETO capability & \multicolumn{4}{c|}{yes}   \\
\hline
\end{tabular}
\end{center}
\end{table}

 Most challenging are high-repetition rate 2D imaging detectors. Therefore, concerning the high frame rate and associated data rate,
 three detector programs are presently working towards producing 2D area detectors 
for the European XFEL. 
Out of these three programs, two will provide detectors optimized for the hard X-ray regime, with optimal performance at photon energies 
above 10~keV: the Large Pixel Detector (LPD) and the Adaptive Gain Integrating Detector (AGIPD).
The DEPFET sensor with Signal Compression (DSSC) is designed for experiments using lower energy X-rays, 
down to a few hundred eV. 
%
%

The LPD detector~\cite{lpd_development_2012}, to be used at the FXE instrument, achieves the front-end required large dynamic range by using three different gain settings 
in parallel, each followed by its own analogue pipeline. 
In order to implement three analogue pipelines with 512 storage cells each in parallel per pixel, the pixel 
size has to be sufficiently larger, e.g. 500~$\mu$m $\times$ 500~$\mu$m.   
The LPD front-end module will include an interposer between the silicon sensor and the ASIC, which gives the 
flexibility to have different pixel sizes and layouts between the silicon sensor and the ASIC. 
%
%
Extra radiation shielding between the sensor and the part of the ASIC, e.g. the memory cells, relaxes the required radiation 
hardness of the ASIC.
The LPD two-tile prototype system was delivered to the XFEL.EU in March 2013 and is under test there.
Dedicated X-ray beam tests at the PETRA III synchrotron at DESY and at the linear accelerator LCLS of the 
prototype are planned for May 2013. 

%
The AGIPD detector~\cite{Henrich2011S11,Shi2010387} consists of a classical hybrid pixel array, with readout ASICs bump-bonded to a silicon sensor. 
The ASIC is designed in 130 nm CMOS technology and uses dynamic gain switching to cover the large dynamic range ($10^4$ photons at $E_{\gamma}=$12~keV), 
and an analogue memory to store recorded images during the 0.6 ms duration of the bunch train. 
The images are subsequently read out and digitized during the 99.4 ms interval between bunch trains.
 The analogue memory consists of two types of storage cells: for amplitude values and for the encoded gain settings. 
The analogue memory is designed to store 352 images, i.e. 352 samples per pixel of size 200 $\mu$m $\times$ 200 $\mu$m. 
To optimize the use of this limited storage depth by overwriting unfit/obsolete images, the memory is operated in random access mode. 
The detector components (sensor and ASIC) are optimized to be radiation tolerant\cite{Schwandt:2012ku,Schwandt:2011pi}.
To verify the performance of the detector, the latest chip assembly (AGIPD0.4) was tested at the P10 beameline of the PETRA III synchrotron 
at DESY\cite{agipd:beamtest}. The required high dynamic range, (10$^4$ photons), was demonstrated and the measured ENC noise was less than 340 e$^-$. 
The results are in agreement with the expected detector performance .
The AGIPD detector is foreseen as the primary detector for the SPB and MID instruments.
%

The DSSC~\cite{porro_development_2012} uses a non-linear response of the active sensor pixels to cope with the large dynamic range, and a digital
 memory to store images inside the pixels. 
An advantage of the DEPFET is the low noise performance, which makes this detector well suited for experiments 
using lower energy X-rays, down to a few hundred eV. 
The DSSC design foresees hexagonal pixels (pixel pitch 204~$\mu$m $\times$ 236~$\mu$m), which give a more homogeneous
 drift field and a faster charge collection than square pixels. 
The DEPFET sensor will be bump bonded to mixed signal readout ASICs. 
The ASICs are designed in 130nm CMOS technology and provides fully parallel readout of the DEPFET pixels. 
The signals coming from the detector, after having been processed by an analogue filter, are immediately digitized and 
locally stored in a custom-designed memory also integrated in the ASICs. 
The advantage of digital storage over analogue is the absence of signal leakage. 
The number of stored frames per macro-bunch is expected to be 640.
This detector is planned to be part of the SQS and SCS instruments.
The first sensors with the final non-linear characteristic will be ready in spring, the full-scale chip submission is planned 
for this summer. 

In parallel to the technology development of detectors, their 
integration into XFEL.EU in terms of mechanics is ongoing. Similarly, work on the calibration strategies and software development
 is progressing. 
A calibration group, which includes representatives from each of the consortia as well as the XFEL.EU 
Detector Development Group members has been set up. 
The goal of the group is to define the detector parameters to be calibrated and the required accuracy,
to develop the strategies to calibrate the different types of detectors, define and build the infrastructure needed for 
calibration and characterization of the different types of detectors, to define calibration and test (re-calibration) procedures, 
to define and implement the calibration data format and to structure and develop and implement calibration software within 
the XFEL.EU software framework (Karabo). More details about the detector test infrastructure will be given in 
Section~\ref{sec:infrastructure}.

In the following subsection we will focus  on the detector geometry requirements driven by the two scientific instruments
where the AGIPD detector will be used as a primary detector, SPB and MID.

\subsection{Detector geometry at the SPB instrument } 
\label{ssec:geometry}
The main scientific cases currently foreseen at the SPB instrument are crystallography of biological nano-crystals (NX) as well as coherent diffractive imaging (CDI) of both  reproducible and non-reproducible particles.

One of the key potentials of CDI lies in its ability to provide quantitative image
contrast, so that each reconstructed real space pixel can be associated with its
sample electron density. 
This imaging modality is possible only if certain requirements of the detection geometry 
are fulfilled. 
As a necessary condition, the diffraction pattern needs to be sampled on a fine enough grid to record its full 
information content. 
In addition, the high-intensity scattering contributions close to the
center of the diffraction pattern need to be measured as completely as possible. 
Any missing information due to the passage of the central beam through the central detector hole, or due to other
blind regions introduced by the modular design of the detectors (non-sensitive areas in between
detector modules) can distort or prevent a quantitative reconstruction~\cite{seibert_single_2011}.
Another demanding requirement in CDI, particularly for larger samples such
as viruses or cells, is the high intensity range of diffraction patterns~\cite{huang_signal--noise_2009}. 
Detectors with a dynamic range of seven orders of magnitude or more are required to cover the full
dynamic range of the diffraction pattern for strongly scattering samples. 
This requirement, however, exceeds the maximum dynamic range of even the newest
integrating detectors suitable for XFEL.EU applications by a few orders of magnitude.
It was shown~\cite{giewekemeyer_detector_2013} for non-crystalline particles that both the mentioned problems, 
i.e. the missing data regions and the high intensity range of certain diffraction patterns, can be mitigated by utilizing
two detector planes: one device close to the sample with a relatively large active area and
central hole (2.5~cm), to measure the high-resolution diffraction data, and one
smaller rear device at about twice or more the distance from the sample to measure
the small-angle diffraction. This second device also needs a central gap or hole to allow for the
passage of the direct beam, which is eventually blocked by a beam dump at the end of
the experimental hutch. The two detectors can operate at different gain settings, by e.g. using filters so that the larger 
dynamic range which was brought up as a requirement is also addressed.
 
Another critical issue is that the resolution in crystallography is geometrically limited by the numerical aperture covered 
by the detector at a certain distance to the sample. If scattering from the sample exceeds the
detector area, the resolution is said to be geometrically limited. 
In a crystal, in contrast to biological single particles, the signal is strongly enhanced 
by the large number of unit cells. 
The crystallographic resolution of an instrument is given by the following equation~\cite{klaus_2013}:
\begin{equation}
\label{eq:diff} 
\Delta x =  \frac{\lambda}{2 \cdot \sin{(\frac{2\theta}{2})}},
\end{equation} 
where $\lambda$ is the wavelength and $2\theta$ is the diffraction angle with respect to the optical axis.
Therefore, the resolution as a function of the detector-sample distance is:
\begin{equation}
\label{eq:diff2}
z =  \frac{D}{2 \cdot \tan{(2 \cdot \arcsin{(\frac{\lambda}{2\Delta x})})}}, 
\end{equation}  
where $D$ is the distance between two opposing outermost pixels on the detector.
Given the feasibility of 1.9~\angstrom~crystallographic resolution for NX at LCLS~\cite{boutet_high-resolution_2012}, a resolution of at least
1.5-2~\angstrom~should  be provided at the XFEL.EU instrument. Taking into account the geometry of the AGIPD detector,  
minimum sample-detector distances of $z_{min} = 13.3-19.4$~cm at a photon energy of 12~keV, and smaller for lower energies 
are required~\cite{klaus_2013}.

\subsection{Angular resolution and pixel size at the MID instrument} 
\label{ssec:lowrep}
One of the most critical parameters for the science cases at the MID instrument is the angular resolution, which
 is determined by the pixel size and the sample-to-detector distance. 
X-ray Photon Correlation Spectroscopy (XPCS), a technique that studies the slow dynamics of various equilibrium and non-equilibrium 
processes occuring in condensed matter systems, is the method which requires the highest angular resolution of all applications. 
A detector with a pixel size of 200$\mu$m $\times$ 200$\mu$m, is not suitable for the 
scientific cases where a high angular resolution is required.
In order to get the highest quality time-correlated data, not more than one speckle per pixel is needed. 
Therefore,  the pixel size should be equal or smaller than the speckle size.
A possible solution can be the use of commercial detectors with sufficient small pixel size but much lower readout rate than 4.5 MHz
which are also being developed to mitigate the risks for day-one operation e.g. the Fast CCD~\cite{fastCCD2009} and 
pnCCD~\cite{struder_large-format_2010}. The Fast CCD  has a limited dynamic range but a small pixel size (30$\times$30 $\mu$m$^2$). 
As the detector was designed for soft X-rays, particular attention has to be paid when running with hard X-ray photons in 
terms of radiation damage. 
In addition, it has to be considered that the quantum efficiency at hard X-ray energies is limited due to the sensor thickness of 200~$\mu$m Si.
The potential for further developing the detector for hard X-ray experiments is being investigated.
Of particular interest would be the use of a thicker sensor increasing the efficiency and obtaining
 a larger dynamic range. 
%

\section{Outlook to other detector projects} 

\subsection{Detectors for spectroscopy and diagnostics} 
\label{ssec:1d}
The unique features of the XFEL.EU, in particular the very high repetition rate, represent a challenge 
for the large-area imaging detectors but also for the smaller-area detectors and make the use of standard commercial 
devices impossible. 
Dedicated solutions are therefore envisaged for small imaging- or strip-detectors.
At the moment, two particular 4.5 MHz small-area detector solutions are under study and are planned to be used at
 the XFEL.EU: a strip detector for hard X-rays and an imaging detector for soft X-rays.
Hard X-rays photon-beam diagnostics as well as hard X-ray absorption and emission spectroscopy at the European XFEL 
make use of strip detectors as detectors for beam spectrometers or as energy-dispersive detectors in combination with 
a crystal as energy-dispersive element. 
The European XFEL is establishing cooperation with the Paul Scherrer Institute in Villigen to develop a new version 
of the Gotthard detector well suited for the XFEL.EU needs. 
Starting from the present detector version, the modifications planned to adapt it to XFEL.EU running 
conditions include the capability of running at an increased frame rate and to provide a veto signal in order to be able to remove non-interesting images. 
In another particular application, resonant inelastic X-ray scattering (RIXS), a Micro-Channel Plate detector matched 
to a delay-line readout is foreseen to be used. 
In this case the European XFEL is planning to use a highly customized solution provided by a German Company. 
 The science-driven detector specifications have been defined and the expected detector performance estimated. 
The detector will be able to record, at 4.5 MHz repetition rate, the photons emitted by the investigated sample, 
expected to be not more than 250 towards the detector with a resolution of the order of ~50 $\mu$m in both directions.

\subsection{Future developments} 

Although the XFEL.EU will start user operation in 2016 and the three main 2D detector projects are expected to cover the need of the 
scientists for the first years of running, plans for the next generation of detectors 
are being made. 
The need for smaller pixel sizes, higher dynamic range  and the possibility of running at 4.5~MHz and to record all the 2700 
pulses of a macro bunch are the main motivations that push the search for new technology solutions. 
The change from 130 to 65 nm CMOS  technology and three-dimensional integrated electronic circuits 
can contribute to this challenge. 
The development of edgeless sensors can help in minimising the inactive detector area. 
Moreover, the need to have high quantum efficiency also at energies greater than  20~keV motivate 
the use materials with higher atomic numbers than silicon for the detectors' active areas.

\section{Detector Test Infrastructure} 
\label{sec:infrastructure}
Accurate calibration and characterization of the different detectors as well as development of 
user-friendly procedures and tools to re-do calibration during the XFEL.EU operation phase 
are very important for the success of the project.  
The detector group at XFEL.EU works on a dedicated test environment for fully assembled detectors, 
detector modules and/or sensor tiles which can be used during the start-up phase of the project and during the operation phase of 
the European XFEL facility. 
The test environment will allow to do  functional tests as well as calibration and performance characterization of 
the detectors in use at the experimental stations of XFEL.EU. 
Since the detector systems foreseen for XFEL.EU consists of components sensitive to surface and particle 
contamination, a clean and controlled infrastructure will be provided for calibration, characterization and testing of 
detectors and detector components. 
Especially for detectors for low energy applications at photon energies below 2 keV and for handling and 
storage of e.g. MCPs, a controlled and clean environment is essential.
For the detector development program of the Detector Group at the XFEL.EU we aim for a calibration, integration, inspection and 
cleaning/wet chemistry laboratory in a clean room in the XFEL.EU headquarter building (XHQ) in Schenefeld. 

As the main laboratory in Schenefeld will not be ready before 2015, the XFEL.EU detector group is temporarily using one of the DESY
experimental halls. The work on developing the infrastructure needed for detector characterization is in progress.

\subsection{X-ray sources for detector calibration and characterization} 
\label{ssec:xray}
To achieve precise results, the characterization and the detailed calibration of the 
2D X-ray cameras, strip and other X-ray sensitive detectors have to be performed using X-ray photons, with characteristics similar to the real conditions at the XFEL.EU experiments. 
The X-ray generators and sources, which will be used for calibration, have to fulfill the following requirements:
\begin{itemize} 
\item 	The time structure of the X-rays has to be similar or as close as possible to 
that of the XFEL.EU. Ideally, short ($\le 50$~ns) X-ray pulses created with a repetition rate of 
up to 4.5~MHz will be required. X-ray sources, which are capable to reproduce the XFEL.EU time structure
(repetition rate, train structure) will be useful to test the timing properties and pulse-to-pulse response 
of the detection systems, e.g. the vetoing of a pre-defined pulse.
\item The X-ray energy of the source shall be similar to that used at the XFEL.EU experiments, 
i.e. 0.5 to 25~keV. This is particularly important for detector performance parameters, which strongly depend on 
the X-ray photon energy. 
Ideally the source should be monochromatic, i.e. with an energy spread significantly smaller than 
the ‘energy resolution’ of the detector. 
\item The intensity of the X-ray beam, which can be defined as:
\begin{equation}
I=  \frac{N_\gamma}{A \cdot \tau}
\end{equation}
where  $N_{\gamma}$ is number of photons, $A$ is the unit area (e.g. pixel size) and $\tau$ is the unit time 
(e.g. pulse length), shall be adjustable and vary from one photon pixel$^{-1}$ pulse$^{-1}$ up to $10^4$ photons 
pixel$^{-1}$ pulse$^{-1}$ at the detector head to be able to calibrate and characterize the detector response over 
a wide dynamic range. Ideally, the X-ray intensity shall be stable in time to a level of  1\% h$^{-1}$ at photon 
intensities $I=10^4$ photons pixel$^{-1}$. In general, the intensity stability of the source has to be better than 
the Poisson noise ($\sqrt{(N_{\gamma} }$).
\item	Illumination of the detector surface:
\begin{itemize}
	\item {\bf Point illumination:} point illumination means that illumination of a single pixel/strip shall be possible. 
 For the actual generation of imaging detectors under development for XFEL, we require an X-ray spot with a diameter 
 of $\le$ 20 $\mu$m. 
	\item {\bf Cluster illumination:} illumination of a cluster of pixels shall be possible to test pixel-to-pixel 
        interaction (e.g. cross talk). Uniform illumination with X-rays with  known emission characteristics on up to 
         16 x 16 pixels is sufficient.
	\item {\bf Uniform flat field illumination:} to study the uniformity of the response of the area detectors uniform 
         flat field illumination shall be possible with homogeneity of the order of 10\% for low spatial frequency
         variations. Ideally the flat field size should be large enough to illuminate an entire active area of a detector 
         or detector module, i.e. approximately 10 cm x 10 cm or larger. Since this is technically difficult or 
         impossible to realize, smaller flat field sizes together with the possibility to move the detector through the 
         X-ray beam are an alternative option.
\end{itemize}
\item The X-ray test setup shall allow testing and operating detector systems under different conditions: 
\begin{itemize}
	\item Vacuum operation: for testing of detectors sensitive to low energy photons, E $\le 1$~keV, 
 a clean environment is required to e.g. avoid surface contamination of the entrance window. Consequently the test setup shall be 
able to be operated at a pressure lower than 10$^{-5}$ mbar. The time to vent and pump the test setup shall be a few hours at maximum 
to reduce overhead times to a minimum. 
	\item Atmospheric pressure operation: for detectors being sensitive to photons with energy above 3 keV, 
              it is sufficient to operate the detector in air or inert gas at ambient pressure.             
\end{itemize}
\end{itemize}
It is very challenging to build an X-ray test environment that meets all the requirements summarized above. 
Therefore, we consider using different sources and test setups which are complementary to each other.
Currently two types of sources are considered for laboratory calibration and characterization of the XFEL.EU detectors: 
radioactive isotopes and X-ray/electron generators.
Long-lived radioactive isotopes are convenient X-ray sources, providing stable intensity 
(half-life of the radioactive isotope $\gg$ time needed for measurement) 
and characteristic emission lines with well-defined photon energies. 
In addition, they are compact, easy to operate and inexpensive in comparison to other X-ray generators.
Radioactive sources can be used for ‘flat-field’ and point illumination if used in combination with a collimator and pinhole mask. 
The achievable intensities are moderate and limited due to the activity of the available isotope sources.
X-ray generators will be used whenever higher X-ray intensities are required and shall provide 
multiple-quanta energies.
The commercially available diffraction or industrial X-ray tubes are mostly operated 
in continuous  mode or at low repetition rates and thus are not sufficient for the 
characterization of the detector parameters which may depend on the repetition rate.
Nevertheless they can be used to calibrate some of the detector parameters 
(e.g. gain conversion). We consider small portable X-ray generators like the Mini-X from AMPTEK Inc..

To be able to characterize the features of the detectors which are mostly connected with the European XFEL unique 
time structure and beam intensity we are planning to use FELs or suitable synchrotrons beams. 
For such tests where the high intensity of the source is not so important,
we are considering to build a custom pulsed multitarget X-ray generator.

The main  subcomponents of the such custom generator are the following:
\begin{itemize}
\item {\bf Electron source:} the electron gun shall provide a pulsed electron beam with electron energies of up to a few tens of keV, 
adjustable in intensity and spot size. We require a beam current adjustable from 1~nA to 20~mA and a spot-size ranging from 
0.1~mm (focused column, spot) to 10~mm (flood beams). The electron gun shall provide pulse widths from approximately 50~ns up to 0.1~ms.
Ideally we require two electron guns, which can be operated in parallel or successively providing two electron pulses with a distance 
equivalent to the XFEL.EU pulse-to-pulse distance of 220~ns.

\item {\bf Multi-Target Anode for X-ray production:} the anode should provide space for approximately ten different target 
materials. It is attached to the shaft of a motor rotating the anode to choose the required target material. 
During operation the anode should stay in a fixed position. 
Due to the high electron currents needed to generate high X-ray intensities, the anode wheel needs to be cooled. 
Most likely a liquid cooling system is required.
\end{itemize}

The X-ray test environment will provide an experimental setup to test and characterize detectors under vacuum and ambient 
conditions with a high intensity pulsed or continuous X-ray beam. 
The test environment will be designed in such a way that several different kinds of X-ray sources can be operated, e.g. 
radioactive isotopes or an X-ray generator. It will consist of several subcomponents  
i.e. a X-ray source, a short beam line of a length of 2 to 3~m and a test chamber for vacuum and ambient operation.

\subsubsection{ Monte Carlo simulation for pulsed multi-target X-ray generator} 
\label{ssec:mc}
An accurate description of the X-ray spectra emitted from targets irradiated by kilo-electron-volt (keV) 
electron beams is of interest for the design and later for the optimization and  
characterization of the X-ray generator which will be built for XFEL.EU detector calibration purposes.
In principle, the different features of X-ray spectra can be computed via the numerical solution of the 
electron-transport equation. 
However, this kind of solution is only possible for relatively simple interaction models and planar geometries. 
Therefore, Monte Carlo (MC) programs are the most suitable methods for the simulation of electron-induced X-ray spectra, 
mostly because they can incorporate realistic interaction cross sections and can be applied to targets with complex geometries.
This is due to the fact that Monte Carlo methods permit to simulate the passage of radiation through matter
taking into account all the relevant physical processes, and all particles (e.g. electrons and photons) can be 
tracked until they stop.   
To optimize the geometry of the generator as well as to estimate the intensity of the X-ray production, a simulation  
based on the general-purpose GEANT4 Monte Carlo tool-kit~\cite{Agostinelli2003250,geant4} was used. 
GEANT4 is a toolkit developed at CERN for the simulation of the passage of particles through matter. 
It is an object-oriented simulation framework which provides a diverse set of software components. 
All aspects of the simulation process, like the geometry of the system, 
 the tracking of the particles through materials as well as the response of sensitive detector components, are included in the simulation.
GEANT4 also provides a set of different models to describe the interaction of particles with matter across a wide energy range.
The standard electromagnetic physics package implements electron, positron, photon and hadron interactions,
but it is not optimized for low energy particles.
Two specific low-energy electromagnetic models are available: the Livermore model, based on the Livermore cross-section and atomic 
data libraries, as well as the Penelope model, based upon the Penelope-code~\cite{penelope}. 
For the present studies the Penelope model was used.

\begin{figure}   
   \begin{center}
   \begin{tabular}{c}
   \includegraphics[height=10cm]{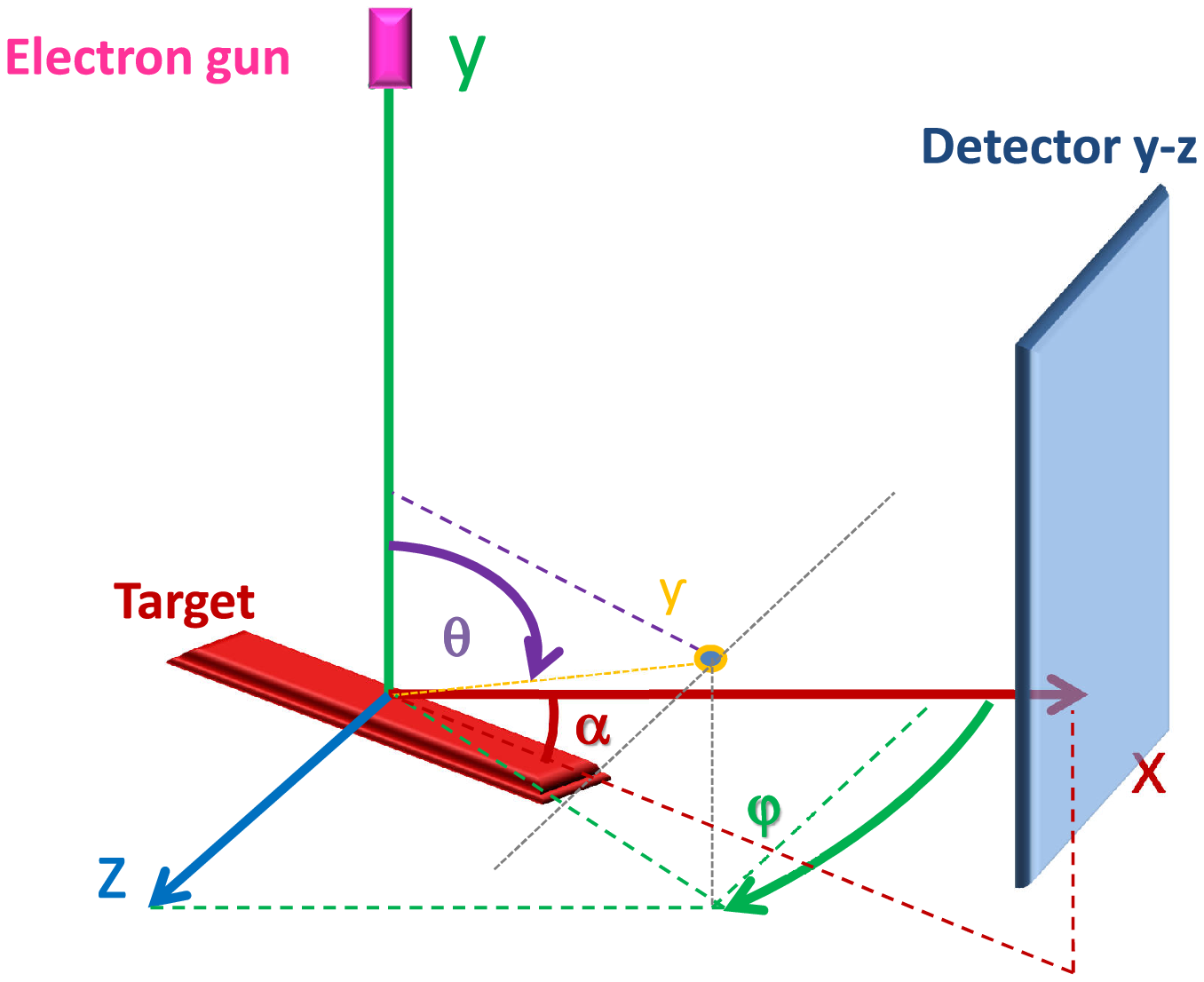}
   \end{tabular}
   \end{center}
   \vspace{-3 cm}
   \caption[energy] 
   { \label{fig:setup} The geometry of the simulated setup.}
   \end{figure}

The geometry of the X-ray generator as implemented in the simulation is shown in Figure~\ref{fig:setup}. 
Mono-energetic electrons with kinetic energies from 30 keV to 100 keV were hitting  
the target. The simulation was done for different target materials, incident angles $\alpha$, 
and electron beam diameters ($d_{beam}$ = 0 - 1~mm). 
Each simulation consists of $4 \times 10^6$ electron.
The procedure of the X-ray production consists in tracking a large number of incident electrons hitting 
the target, until they are absorbed or emitted from it, and calculating the amount of Bremsstrahlung and characteristic
 photons produced during the electrons' traverse of the target.
The procedure starts with the definition of the electron source with the energy $E_e$. 
When the electrons hit the target, the code transports them inside the target material, producing secondary photons 
through Bremsstrahlung and ionization, until they are stopped. 

\begin{figure}
   \begin{center}
   \begin{tabular}{c}
   \includegraphics[height=7cm]{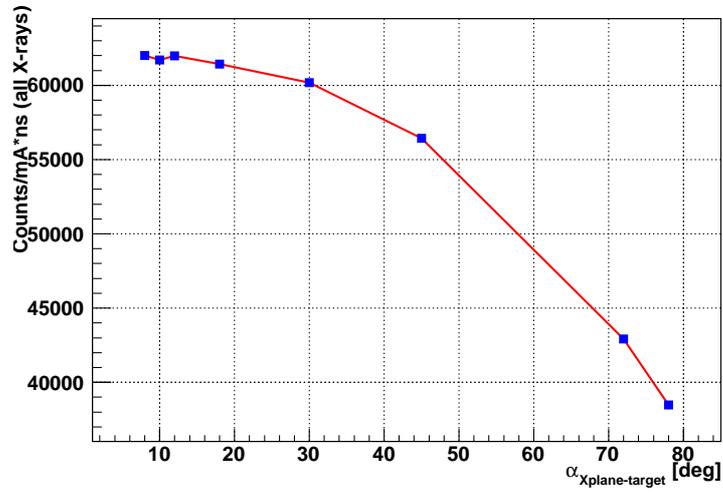}
   \end{tabular}
   \end{center}
   \caption[energy] 
   { \label{fig:target} 
    Number of X-ray quanta as a function of the incident angle $\alpha$ for the electrons with energy $E_e=100$~keV incident on 5~mm thick Cu target.}
   \end{figure}

The number of the X-ray quanta as a function of the incident angle $\alpha$ for the electrons with  energy $E_e$=100~keV incident 
on a 5~mm thick Cu target is presented in Figure~\ref{fig:target}. 
The maximum intensity of the X-ray beam is achieved for such geometries where the angle between the target
and $x$-axis is smaller than 30$^{\circ}$. 

\begin{figure}   
   \begin{center}
   \begin{tabular}{cc}
   \includegraphics[height=5.5cm]{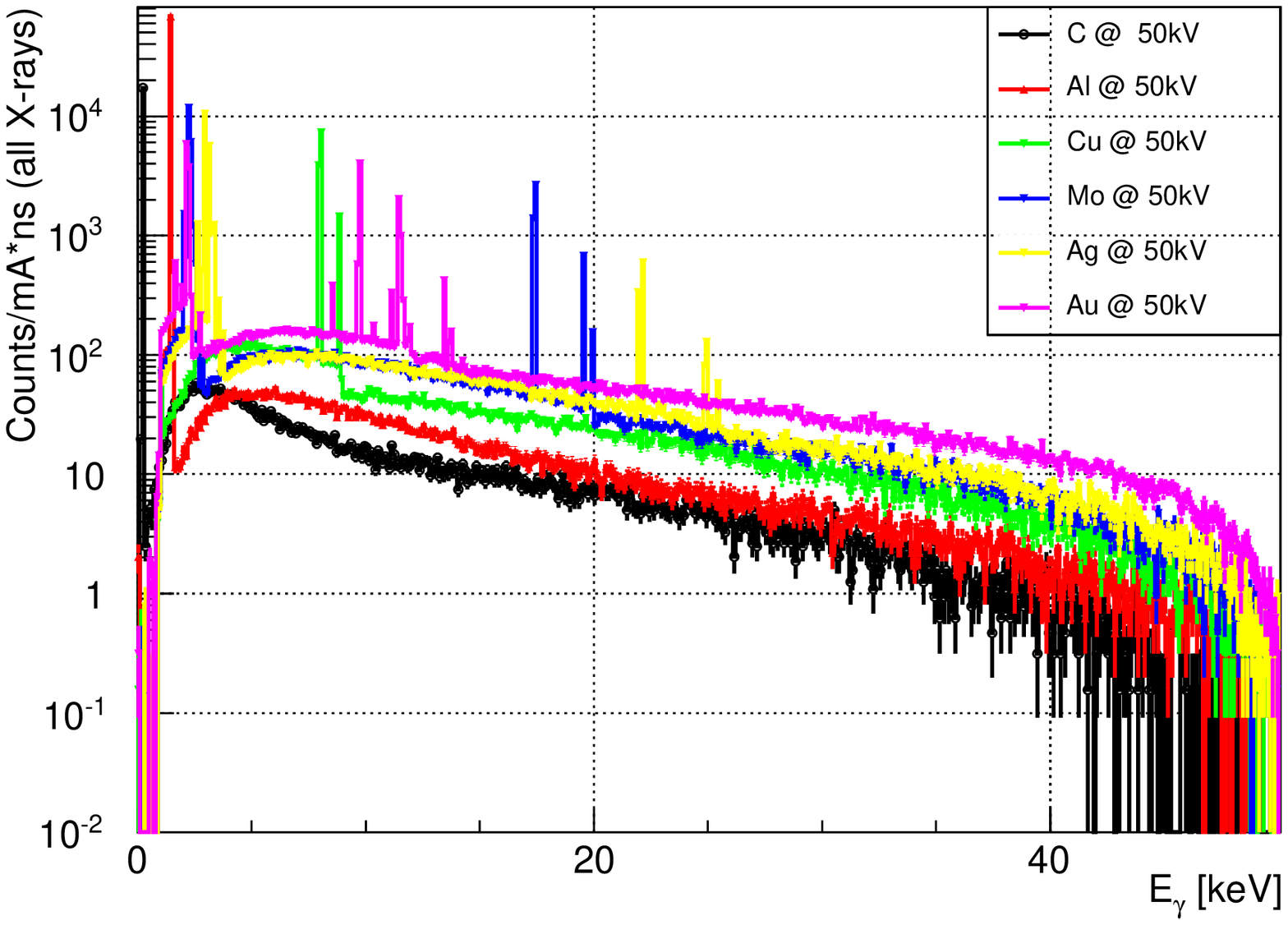}& 
   \includegraphics[height=5.5cm]{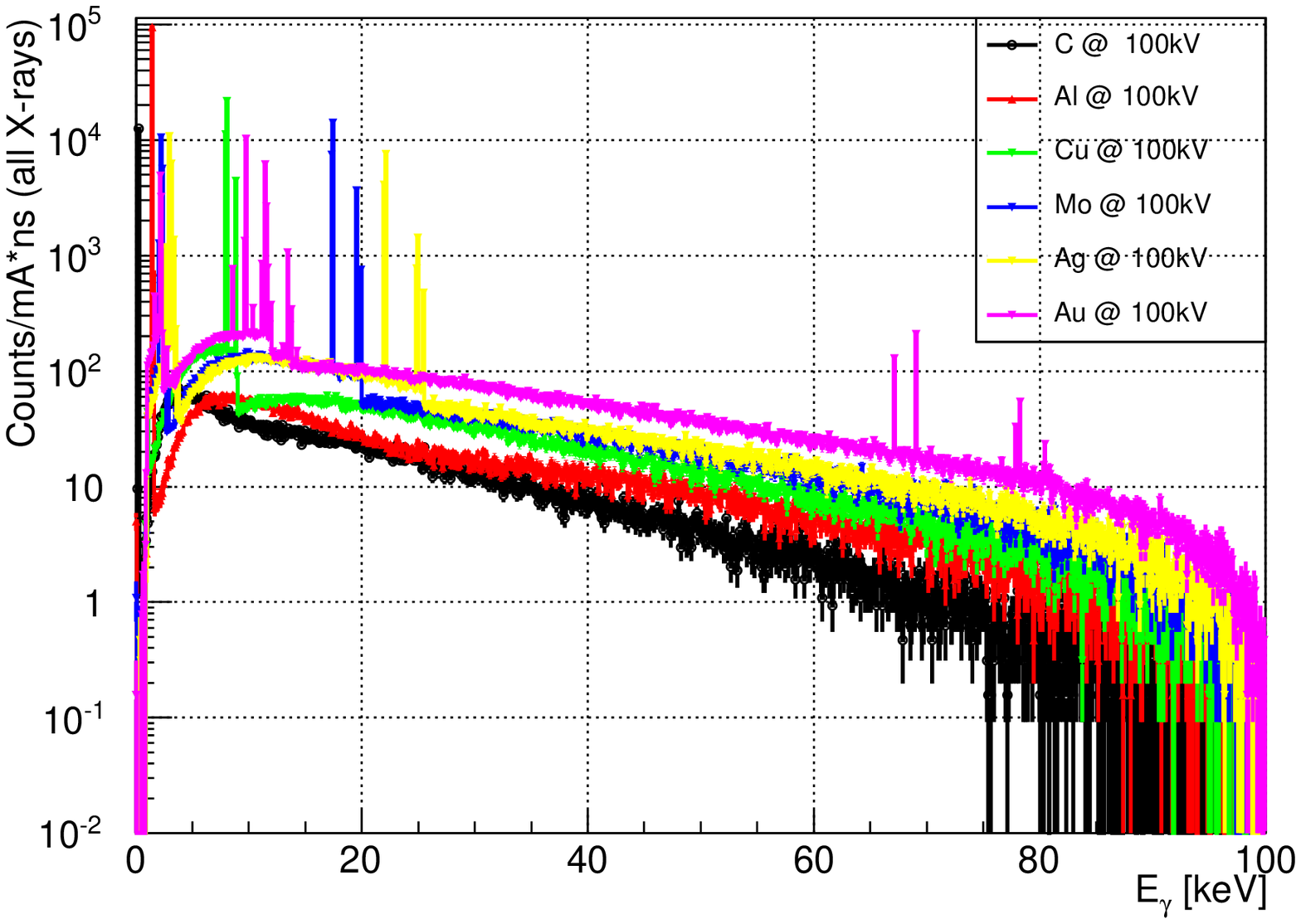} 
   \end{tabular}
   \end{center}
   \caption[energy] 
   { \label{fig:spectra} 
    The X-ray spectra produced by the electrons with energy $E_e$=50~keV (left plot) and 100~keV (right plot) hit on the different targets at the 
 incident angle $\alpha = 18^\circ$.}
   \end{figure}

The X-ray spectra of different target materials for electron beam energy, $E_e=$ 50 and 100~keV, 
are shown in Figure~\ref{fig:spectra}. The results are normalized
to the operation conditions corresponding to an electron beam current of $I_e=$1~mA,
delivered in $\Delta t=$1~ns. 
The characteristic lines as well as the continuos spectra
(Bremsstrahlung) are presented. 
 
\begin{figure} 
   \begin{center}
   \begin{tabular}{c}
   \includegraphics[height=5.5cm]{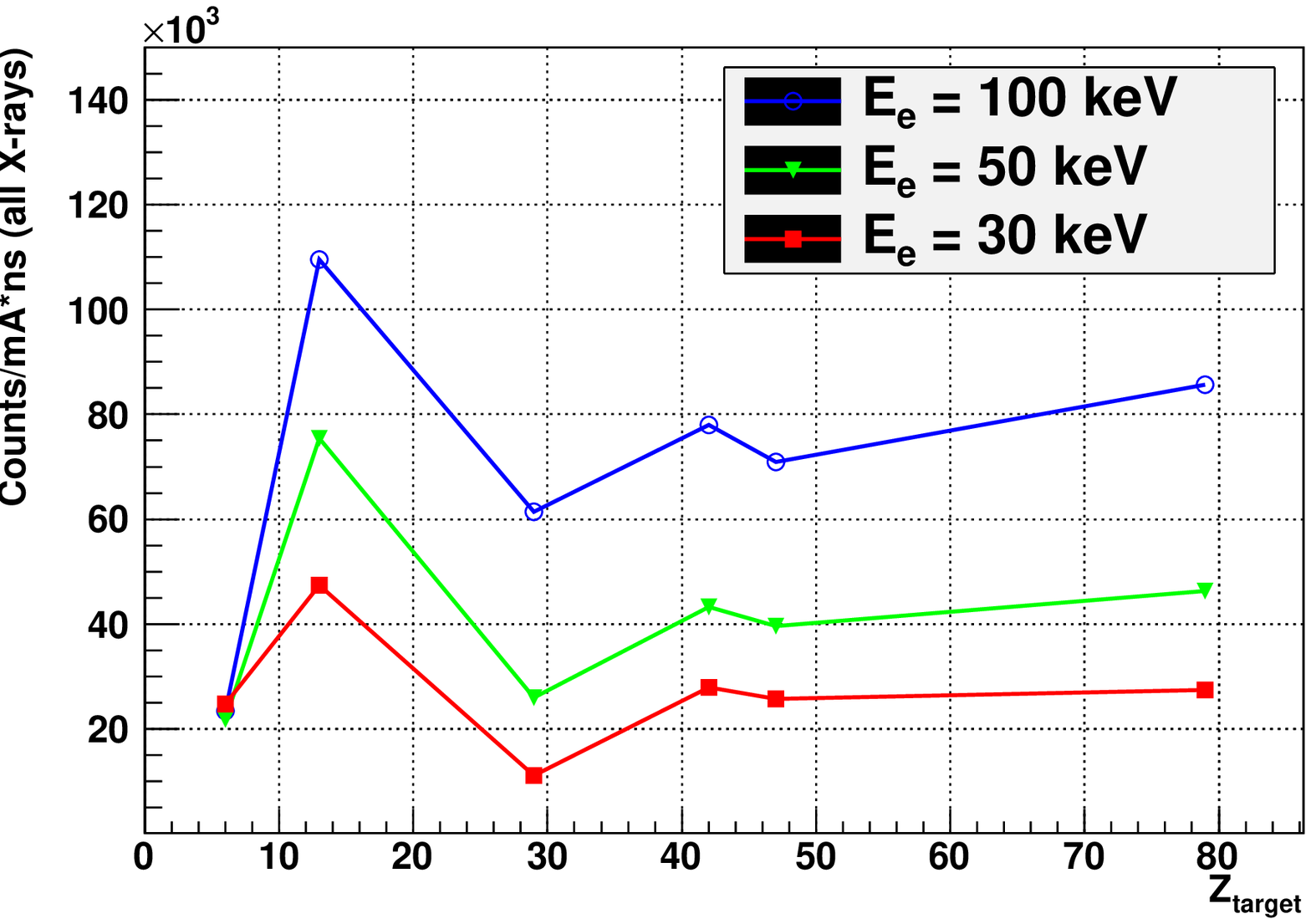}
   \includegraphics[height=5.5cm]{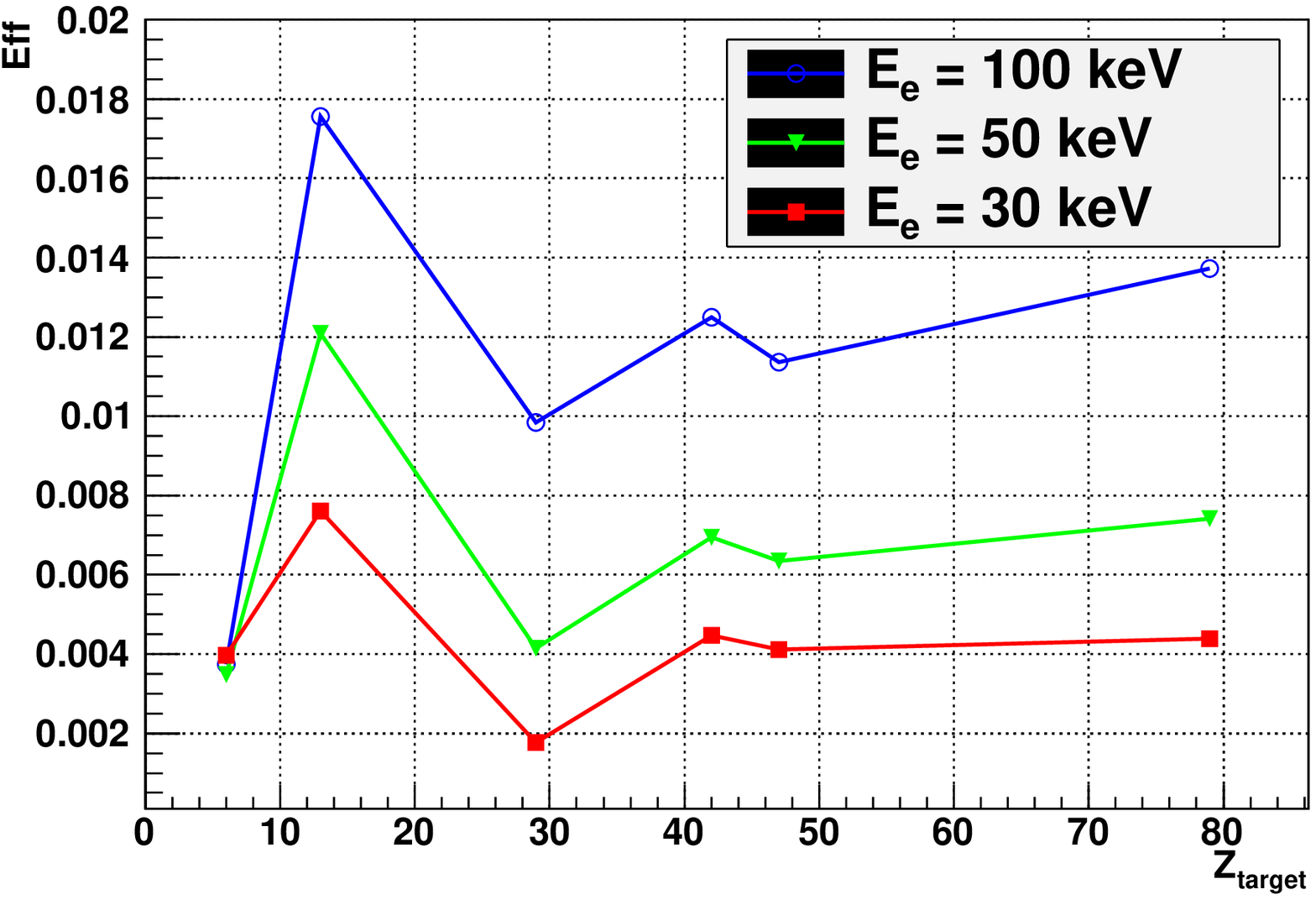}
   \end{tabular}
   \end{center}
   \caption[energy] 
   { \label{fig:ngamma} 
    Number of X-ray quanta as a function of the atomic number of the target material (left plot) 
    produced by the electrons with energies $E_e$=30, 50 and 100~keV hit on the target at the incident angle $\alpha = 18^\circ$ and 
    the efficiency of the X-ray production (right plot).}
   \end{figure}

The number of X-ray photons produced in $\Delta t=$ 1~ns as a function of the atomic 
number of the target material, $Z$, for the different electron energies $E_e$=30, 50 and 100~keV and $I_e$= 1~mA
as well as efficiency of the X-ray production are shown in Figure~\ref{fig:ngamma}. The efficiency for X-ray production
is varied from 0.4 up to 1.8~\% for electron energy $E_e= 100$~keV and respectively lower for the lower energies.

The results presented in Figures~\ref{fig:target}~-~\ref{fig:spectra} refer to the full angular range of the produced X-rays.
%
Assuming the electron energy of $E_e$ = 50~keV, the electron current of $I_e$ = 20~mA and the pulse length of 50~ns, the number of 
photons per pulse in the detector plane of size 20 $\times$ 20~cm$^2$  at 20 cm distance from the source is approx.~2~$\times$~10$^6$.
This yields two hits per pixel of size 200 $\times$ 200 $\mu$m$^2$.  
The number of photons per pixel can be increased by using X-ray focussing optics (e.g. policapillary optics).
Polycapillary focusing optics collect a large solid angle of X-rays from an X-ray source and focus them to a spot as small as a tens of $\mu$m. 
The resulting X-ray flux density obtained is a few orders of magnitude higher than that obtained with a conventional pinhole collimator. 
A system design is under evaluation.
Taking in to account the presented results, an electron source operated in pulsed mode with energies up to 50~keV
and electron current of 20~mA would be sufficient for the laboratory detector characterisation. 
The incident angle should not be larger than 30$^\circ$ and the beam diameter shall range from 0.1 to 1~mm.

\section{Software Development at the European XFEL} 
\label{sec:sw}
The XFEL.EU is developing a dedicated software framework, Karabo~\cite{karabo_2013}, which integrates both controlling and data-analysis
within a single design. 
Karabo is being written in both C++ and Python. 
It is a distributed design, with messages being passed either via a broker or via peer-to-peer (p2p) connections over a network. 
Its core are transparently distributable servers, which aggregate functionality via hot-plugable ''devices''. 
A device can be a control device, e.g. for steering a motor, a monitoring device, e.g. for measuring a pressure, or 
a computational device, e.g. for data processing. 
Interactivity is provided either from the command line interface (CLI) using IPython 
with automatic command-completion or from a PyQt-based graphical user interface (GUI) which aggregates (time-evolving) parameters, control and 
data-flows as well as state-, logging- and error-information. 
Within this GUI complex control and processing graphs can be laid out allowing for the design of e.g. data-analysis pipelines.
 
\subsection{Detector-specific software in Karabo framework} 
\label{ssec:karaboo}
Within the Karabo framework there will be two flavors of detector specific software: 
control software and analysis/calibration software.
\begin{itemize} 
\item {\bf Control software:} includes software which is needed for controlling and monitoring the detector and closely related components.
This includes e.g. sensor voltages, front-end electronic timings, cooling parameters, vacuum-pressures and corresponding safety-interlocks
 or mechanical positioning of the detector. Such software must necessarily be state-aware, i.e certain commands are only available in 
certain hardware contexts/modes.
\item {\bf Calibration software:} includes software which is used to apply calibration procedures to detector data resulting in data-sets
which can be used for scientific analysis. Different calibration steps will generally be realized as individual software components (devices)
which can be arranged into a processing pipeline. Calibration software also includes the analysis methods necessary for obtaining calibration parameters. 
Especially the application of calibration should be handled near-online, in order to provide scientists with a rapid feedback
on their instrument and experiment.
\end{itemize}

\subsection{Simulation }  
\label{ssec:sim}
Computer-based simulations play an important role for an in-depth understanding of detector performance and characteristics 
such as the achievable dynamic range, the minimum noise or radiation induced performance deterioration. 
The required simulation tools are either inhouse developments, or derived from software used by the detector consortia during development.

In addition to estimating detector characteristics, Monte Carlo simulations are being used to optimize calibration hardware and procedures 
(e.g. X-ray tube simulation in Section~\ref{ssec:mc}). Furthermore, an end-to-end simulation is planned in the long term. 
This simulation aims to model as many aspects of the European XFEL as possible, starting from the electron source and ending with ready-to-analyze,
simulated detector data of a simulated sample.

\section{Conclusion} 
\label{sec:conclusion}
Detector development at the XFEL.EU for fast (4.5 MHz) imaging detectors is supported by three external detector consortia. 
The first two-tile LPD system was delivered to the XFEL.EU for laboratory tests and characterization. 
The system will be tested on the PETRAIII beam at DESY in May.
A prototype of the AGIPD detector is foreseen at the begining of 2014. 
In addition, as a backup solution and day-one-option two projects for low-repetition rate  detectors are ongoing: pnCCD and FastCCD.  
Two additional detector activities have also started recently: 1D detector program for spectroscopy and diagnostics as well as
small area 2D detectors.

In parallel to the detector development, the XFEL.EU detector group is building up the laboratory  infrastructure needed for 
the detector calibration and characterization  and develops the software for the data analysis and detector calibration and characterisation.



\acknowledgments     
 
We thank all the members of the detector consortia, DSSC, AGIPD and LPD for their effort in the XFEL.EU detector program and for the 
effective exchange of information. 
We thank the PSI detector group, Bernd Schmitt and Aldo Mozzanica in particular,  
and Andreas \"Olsner, from Surface Concept, for the proficient communication.


\bibliography{spie_proc}   
\bibliographystyle{spiebib}   

\end{document}